\documentclass[epsf,useAMS,usenatbib]{mn2e}

\usepackage{lscape}
\usepackage{graphicx}
\usepackage{natbib}
\usepackage{graphics}

\def \vhel{\ifmmode{~V_{{\rm HEL}}}\else{~$V_{{\rm HEL}}$}\fi}
\def \vsys{\ifmmode{~V_{{\rm SYS}}}\else{~$V_{{\rm SYS}}$}\fi}
\def \HA {\ifmmode{{\rm\H}\alpha}\else{${\rm\ H}\alpha$}\fi}

\def \msun{\ifmmode{{\rm\ M}_\odot}\else{${\rm\ M}_\odot$}\fi}

\def \myr{\ifmmode{{\rm\ M}_\odot{\rm\ yr}^{-1}}
         \else{${\rm\ M}_\odot$ yr$^{-1}$}\fi}

\def \mdot{\ifmmode{\dot{M}}\else{$\dot{M}$}\fi}
\def \tena#1 #2 {\ifmmode{#1 \times 10^{#2}}\else{$#1 \times 10^{#2}$}\fi}
\def \kms{\ifmmode{~{\rm km\,s}^{-1}}\else{~km s$^{-1}$}\fi}

\title[Impact of tangled magnetic fields on AGN-blown bubbles]{
Impact of tangled magnetic fields on AGN-blown bubbles}
\author[M. Ruszkowski et al.]{M. Ruszkowski,$^1$\thanks{E-mail:
mr@mpa-garching.mpg.de (MR)} T.A. En{\ss}lin$^1$, M. Br{\"u}ggen$^2$, S. Heinz$^3$,
\& C. Pfrommer$^4$ \\
\\
$^1$Max Planck Institute for Astrophysics, Karl-Schwarzschild-Str. 1,
85741 Garching, Germany\\
$^2$International University Bremen, Campus Ring 1, Bremen, Germany\\
$^3$Department of Astronomy, University of Wisconsin, 475 N Charter Street 
Madison, WI 53706, USA\\
$^4$Canadian Institute for Theoretical Astrophysics, 60 St. George Street
Toronto, Ontario, M5S 3H8, Canada
}
\begin{document}

\date{Submitted 2006 December}

\pagerange{\pageref{firstpage}--\pageref{lastpage}} \pubyear{2006}
\maketitle
\label{firstpage}

\begin{abstract}
There is growing consensus that feedback from 
active galactic nuclei (AGN) is the main
mechanism responsible for stopping cooling flows in clusters of galaxies.
AGN are known to inflate buoyant bubbles that supply mechanical power
to the intracluster gas (ICM). High Reynolds number hydrodynamical simulations 
show that such bubbles get entirely disrupted within 100 Myr, 
as they rise in cluster
atmospheres, which is contrary to observations. This artificial mixing
has consequences
for models trying to quantify the amount of heating and star formation
in cool core clusters of galaxies. It has been suggested
that magnetic fields can stabilize bubbles against disruption. 
We perform magnetohydrodynamical (MHD) simulations of fossil bubbles
in the presence of tangled
magnetic fields using the high order {\it PENCIL} code. 
We focus on the physically-motivated case where thermal pressure dominates
over magnetic pressure and 
consider randomly oriented fields with and without maximum helicity
and a case where large scale external fields drape the bubble.
We find that helicity has some stabilizing effect.
However, unless the coherence length of magnetic fields exceeds the
bubble size, the bubbles are quickly shredded. 
As observations of Hydra A suggest that lengthscale of magnetic 
fields may be smaller then typical bubble size,
this may suggest that other mechanisms, such as viscosity,
may be responsible for stabilizing the bubbles.
However, since Faraday rotation observations of radio lobes 
do not constrain large scale ICM fields well if they are aligned with
the bubble surface, the draping case may be a viable
alternative solution to the problem.
A generic feature found in our simulations is the formation of 
magnetic wakes where fields are ordered and amplified. We suggest that
this effect could prevent evaporation by thermal conduction 
of cold H$\alpha$ filaments observed in the Perseus cluster.
\end{abstract}


\section{Introduction}

Active galactic nuclei play a central role in explaining the
riddle of cool core clusters of galaxies. One of the unsolved problems
of AGN feedback in clusters is the issue of morphology and stability of buoyant
bubbles inflated by AGN and the efficiency of their mixing with the surrounding
ICM. 
It is important to understand
the process of bubble fragmentation and its eventual mixing with the
rest of the ICM in order to quantify mass deposition
and star formation rates in cool cores clusters.
In the best-studied case of the Perseus cluster (Fabian et al. 2006)
observations indicate that such bubbles
can remain stable even far from cluster centers where they were created
(see cup-shaped feature northwest of the center of the
cluster or a similar yet tentative feature to the south in their
Figure 3).
One possible explanation for this phenomenon is that the intracluster
medium is viscous and that viscosity suppresses Rayleigh-Taylor
and Kelvin-Helmholz instabilities on their surfaces.
Reynolds et al. (2005) performed a series of numerical
experiments, compared inviscid and viscous cases and quantified how
much viscosity is needed to prevent bubble disruption. For viscosity
at the level of 25\% of the Braginskii value they obtained results
consistent with observations. The same problem has been considered
analytically by Kaiser et al. (2005) who computed the instability growth
rate as a function of scale and viscosity coefficient. An alternative
possibility is that bubbles are made more stable by 
significant deceleration during their initial evolution 
(Pizzolato \& Soker 2006). 
The ``viscous solution'' is very appealing as it also
addresses the issue of dissipation of sound waves, weak shocks
as well as $g-$modes (Fabian et al. 2003a). However, 
it is not entirely clear what the level and nature of
viscosity in the ICM really is. This is because the ICM 
and the bubbles themselves are known to be
magnetized and magnetic fields may 
reduce transport processes (but see 
Lazarian 2006 and Schekochihin \& Cowley 2006). 
Although magnetic fields in clusters are known to have
plasma $\beta > 1$,  ($\beta \equiv P_{\rm gas}/(B^2/2\mu_{o}$); e.g.,
Blanton et al. 2003) 
they may in principle have a
strong effect on suppressing Kelvin-Helmholz and Rayleigh-Taylor
instabilities. Jones \& De Young (2005) considered 
the evolution of bubbles in a magnetized ICM 
by performing two-dimensional numerical MHD simulations 
and found that
bubbles could be prevented from shredding even when $\beta$ is as high
as $\sim 120$. A somewhat different conclusion was reached by 
Robinson et al. (2004) who simulated magnetized bubbles with the {\it
FLASH} code (Fryxell et al. 2000) in two dimensions and found out
that a dynamically important magnetic field ($\beta < 1$) is required
to maintain bubble integrity.
Both of the above-mentioned simulations were performed in 2D and for very
idealized magnetic field configurations such as, 
for example, doughnut-shaped fields
with symmetry axis parallel to the direction of motion. 
Recently, Nakamura, Li, \& Li (2006) considered the stability of Poyinting-flux
dominated jets in cluster environment.
Here we focus on a later stage in the outflow evolution and consider 
physically-motivated magnetic field case of
$\beta > 1$ and 
more realistic (stochastically tangled) field configurations in fossil bubbles
(i.e., after the transition from the momentum-driven to
buoyancy-driven stage) than in previous buoyant stage MHD simulations.
We show that only when the power spectrum cutoff
of magnetic field fluctuations is larger than the bubble size can the
bubble shredding be suppressed. 
It is possible that such ``draping'' 
case is not representative of typical cool core fields 
(Vogt \& En{\ss}lin 2005) and, if so, other mechanism, such as viscosity, 
would be required to keep the bubbles stable. 
However, Vogt \& En{\ss}lin (2005) estimated magnetic power spectra 
in Hydra A only which shows extraordinarily strong AGN outbursts.
We also note that such  
Faraday rotation observations of radio lobes are only weakly sensitive
to large scale magnetic fields aligned with the bubble surface
and rely on certain untested assumptions. Therefore, we argue that the draping 
case may be a viable solution to the problem of bubble stability.\\
\indent
In Section 2 we present the simulation setup and 
the justification for the parameter choices. Section 3 presents
results and Section 4 our conclusions. The 
Appendix discusses in detail our method
of setting up initial magnetic field configurations.

\section{Simulation details}
\subsection{The code}
The simulations were performed with the {\it PENCIL} code. (Dobler et
al. 2003, Haugen et al. 2003, Brandenburg et al. 2004, Haugen et al. 2004)
Although {\it PENCIL} is non-conservative, it is a highly accurate 
grid code that is sixth order in space and third order in time.
It is particularly suited for weakly compressible
turbulent MHD flows.
Magnetic fields are implemented in terms of a vector
potential so the field remains solenoidal throughout simulation
(i.e., no divergence cleaning of the magnetic field is necessary).
The code is memory efficient, uses Message-Passing
Interface and is highly parallel. The {\it PENCIL} code is better suited 
for the simulations of stability of magnetized bubbles than 
smoothed particle magneto-hydrodynamical codes as it 
can better capture certain aspects of the bubble (magnetohydro) dynamics,
such as Kelvin-Helmholtz instability, even for high density
contrasts.
and suffers less from numerical diffusion of the magnetic field.
Although no formal code comparison has been made, 
other well known 
MHD mesh codes, such as {\it FLASH} or {\it ZEUS}, likely require
comparatively higher resolution to achieve the same level of magnetic flux 
conservation as the {\it PENCIL} code.

\subsection{Initial conditions}

\subsubsection{Density and temperature profiles}
The cluster gas was assumed to be isothermal with temperature equal to
10 keV (this temperature was used to minimize the ratio of code viscosity to 
the Braginskii value; see Section 2.2.4 for more explanation). 
The gas was initially in hydrostatic equilibrium and subject to 
gravitational acceleration given by a sum of two isothermal potentials

\begin{equation}
g(r) =-\frac{2\sigma_{a}^{2}}{(r+r_{a})}   
      -\frac{2\sigma_{b}^{2}}{(r+r_{b})},
\end{equation}

\noindent
where $\sigma_{a}=1.41$, $\sigma_{b}=2.69$, $r_{a}=30.0$ and
$r_{b}=5.0$ in code units (see below) were the parameters were 
chosen to give a pressure profile consistent with that observed in
clusters. Central gas density was $10^{-25}$ g cm$^{-3}$
(i.e., about $5.2\times 10^{-2}$ electrons per cm$^{-3}$).
Self-gravity of the gas was neglected.
The profiles of pressure and temperature are shown in Figure 1.
We consider ideal gas with adiabatic equation of state with $\gamma =5/3$.
The bubbles were underdense by a factor
of ten with respect to the local ICM and its temperature was 
increased by the same factor to keep it in pressure equilibrium with the
surrounding gas. The remaining bubble parameters are mentioned in
section 2.2.4 where we discuss code units. The 
pressure in the ICM changes significantly over the height of the bubble. It
is for this reason that we modified the density and temperature on a
"point-by-point" basis. That is, at every location within the bubble,
we increased the temperature by a constant factor and decreased the density by
the same factor while keeping the pressure at the same level as the
pressure at the same distance from the cluster center away from the
bubble. This way, the initial pressure distribution is smooth and
there is no strong departure from
perfect equilibrium. 
Although the density contrast of the bubble is relatively low, this
has minimal effect on our results. This is
because the buoyancy velocity is proportional to $( 1 -
\rho_{\rm bubble}/\rho_{\rm icm} )^{1/2}$ which is insensitive to 
$\rho_{\rm bubble}$ as long as it is much
smaller than $\rho_{\rm icm}$.  
The reason we opted for such relatively low density contrast
is numerical: higher bubble temperatures would have required smaller
timesteps to achieve numerical stability.
We also note that some real
bubbles may have smaller density contrasts. A well-known extreme example 
is the Virgo cluster where buoyant bubbles in
the radio show very little corresponding structure in the X-ray
maps.

\subsubsection{Magnetic fields}
We consider magnetic fields inside the bubbles and in the ICM 
that are dynamically unimportant in the
sense that their plasma $\beta$ parameter is greater than unity.
That is, magnetic pressure may become important when compared to 
the bubble ram pressure associated with the gas motions in the ICM but
it is generally small compared to the pressure of the ICM in our
simulations. Analysis of cluster bubbles  by Dunn et al. suggest that 
$\beta^{-1}$ is roughly in the range (10$^{-3}$, 0.3) with the mean
$\langle\beta^{-1}\rangle =0.06$ 
and median of 0.03 (see sample of Dunn et al. 2004, Dunn et al. 2005 and 
Dunn 2006, Dunn priv. communication). 
Other evidence for high $\beta$ comes from observations of Faraday
rotation measure and the lack of Faraday depolarization in bright
X-ray shells in Abell 2052 as observed by Blanton et al. (2003). They infer
$\beta\sim 30$. These arguments motivated us to consider high $\beta$ cases. 
However, we note that low-$\beta$ is not ruled out beyond reasonable 
doubt by the above observations and theoretical considerations and, thus, 
its consequences for bubble dynamics should be investigated further.\\
\indent
Regarding the geometry of magnetic fields, we consider
random isotropic fields with coherence length smaller than the bubble
sizes (hereafter termed ``random''), 
isotropic helical fields with 
coherence length smaller then the bubble size (termed ``helical''
below; helicity is defined as $\int {\bf A} \cdot {\bf B}dV$, where 
${\bf A}$ and ${\bf B}$ are vector potential and magnetic field 
respectively), and 
the ``draping'' case of isotropic fields characterized by coherence
length exceeding bubble size as well as a non-magnetic case.\\
\indent
Magnetic draping has been considered previously 
in the context of merging cluster cores and radio bubbles 
using analytical approach by Lyutikov (2006). He found that even 
when magnetic fields are dynamically unimportant throughout the ICM,
thin layer of dynamically important fields can form around merging 
dense substructure clumps (``bullets'') and prevent their disruption.
We suggest that, depending on the unknown value of magnetic diffusivity,
there may be some relic magnetic power spectrum 
with a smaller amplitude than the freshly injected one (either by
AGN bubbles or dynamo-driven) that extends to scales larger than the 
bubble size and that provides draping fields to stabilize the bubbles.\\
\indent
The helical case is motivated by the fact that magnetic helicity is
conserved for ideal MHD case.
Moreover, helicity is proportional to the product of
magnetic energy and typical lengthscale and, thus, fragmentation is
not energetically preferred in the sense that it would lead to local
increase of magnetic energy. Another justification for helical fields
is that they may be responsible for explaining circular polarization
of certain radio sources, and especially its sign persistency
(En{\ss}lin 2003 but see Ruszkowski \& Begelman 2002).
It is conceivable that magnetic helicity that could produce such signal
in jets could survive till the buoyant stage.\\
\indent
The ``random'' case is motivated by 
the work of Vogt \& En{\ss}lin (2005) who estimate power spectrum of
magnetic field fluctuations in Hydra A
from Faraday rotation maps. We are conservative in our
choice of parameters in the sense that we use fields of somewhat higher 
coherence length that  are more likely to stabilize bubbles. \\

\subsubsection{Vector potential setup}
In setting up initial magnetic field configuration we 
ensured that the following conditions are met:\\

\begin{itemize}

\item magnetic fields must be divergence-free
\item bubble and environmental fields must be isolated, i.e., no magnetic field
lines should penetrate the surface of the bubble
\item magnetic fields must be characterized by the required power spectra
\item when necessary, magnetic helicity may be imposed 
\item magnetic plasma $\beta$ parameter of the bubble and the ICM 
must be independently controllable
\item all considered 
field configurations must 
result (after suitable modifications) 
from the same white noise Gaussian random numbers. This ensures that 
the differences in evolution are entirely due to model parameters 
and not due to different random realizations.

\end{itemize}

\indent
The details of the algorithm used to set up the initial magnetic field
configuration is presented in the Appendix..

\subsubsection{Code units and resolution}
We consider the the following code units: one length unit corresponds
to 1 kpc, one velocity unit to 3$\times 10^7$ cm/s and one density
unit to 10$^{-24}$ g/cm$^3$. This corresponds to one time unit of
approximately 3.3$\times 10^6$ years. The box size is 100 code length
units on a
side and shock-absorbing boundary conditions were used.
This choice of boundary conditions was dictated by the fact that 
we wanted to exclude the Richmyer-Meshkov instability due to reflected
waves passing through the bubble as a potential destabilizing mechanism.
The bubble size was $d=25$ in diameter and 
its center was offset by 20 code length units from the cluster origin.
The bubble was underdense by a factor
of ten with respect to the local ICM and its temperature was 
increased by the same factor to keep it in pressure equilibrium with the
surrounding gas. Our grid size is 200$^3$ zones.
The {\it PENCIL} code requires some viscosity and resistivity to run
properly. Minimum required amount of physical viscosity that the 
code needs to prevent numerical instabilities from developing 
decreases as the numerical resolution is increased.
We use viscosity $\nu = 0.07$ and magnetic resistivity
$\eta = 0.07$. For typical values of electron density 
($\sim 0.01$ cm$^{-3}$) and temperature (10 keV) in our simulations
this value of viscosity corresponds to 0.014 of the
Braginskii value. We note that for lower gas temperatures this ratio
would be higher which motivated our choice of temperature.
For the parameters considered here, in the initial
stages in our simulations, 
maximum velocities in code units are 
roughly $u_{\rm max}\sim 6$. 
Note that for the adopted parameters the sound speed in the
ICM is $c_{s}=5.44$ but the gas inside the bubbles is much hotter and
so the actual gas velocities can be higher then $u_{\rm max}$ without
``violating'' the sound speed limit. For the bubble size $d=25$ 
and gas velocity $u_{\rm max}=6.0$, 
hydrodynamical and magnetic Reynolds numbers are 
roughly $Re=u_{\rm max}d/\nu\sim 2000$.
Such values clearly lead to quick development of Rayleigh-Taylor
and Kelvin-Helmholtz
instabilities for unmagnetized bubbles.

\section{Results}

\begin{figure}
\centering
\rotatebox{0}{\mbox{\resizebox{8.6cm}{!}{\includegraphics{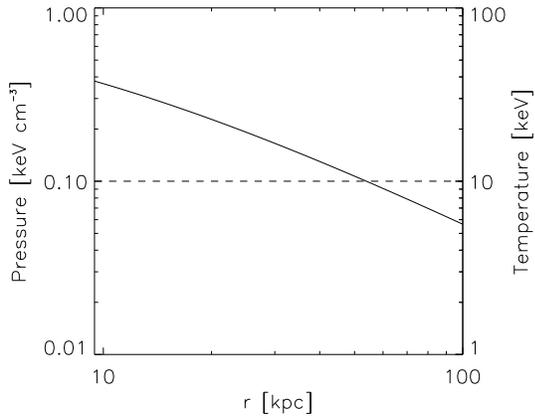}}}}
\caption{\label{}{Initial pressure (solid line) and
temperature profiles.}}
\end{figure}

\begin{figure}
\centering
\rotatebox{0}{\mbox{\resizebox{8.6cm}{!}{\includegraphics{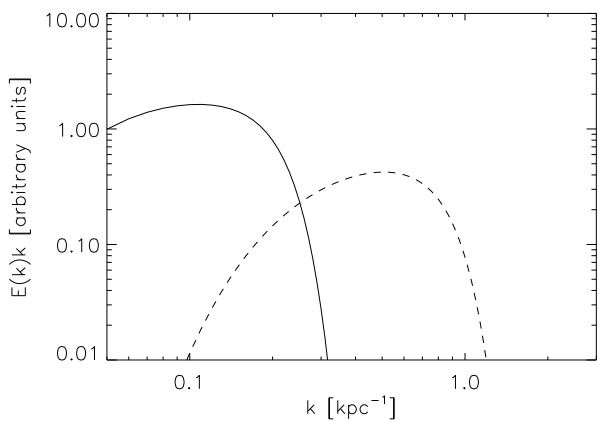}}}}
\caption{\label{}{One dimensional energy 
power spectra of initial magnetic field
fluctuations. Solid line shows energy spectrum corresponding to our
``draping''. Dashed line is for the ``random'' case.}}
\end{figure}

\begin{figure}
\centering
\rotatebox{0}{\mbox{\resizebox{9.0cm}{!}{\includegraphics{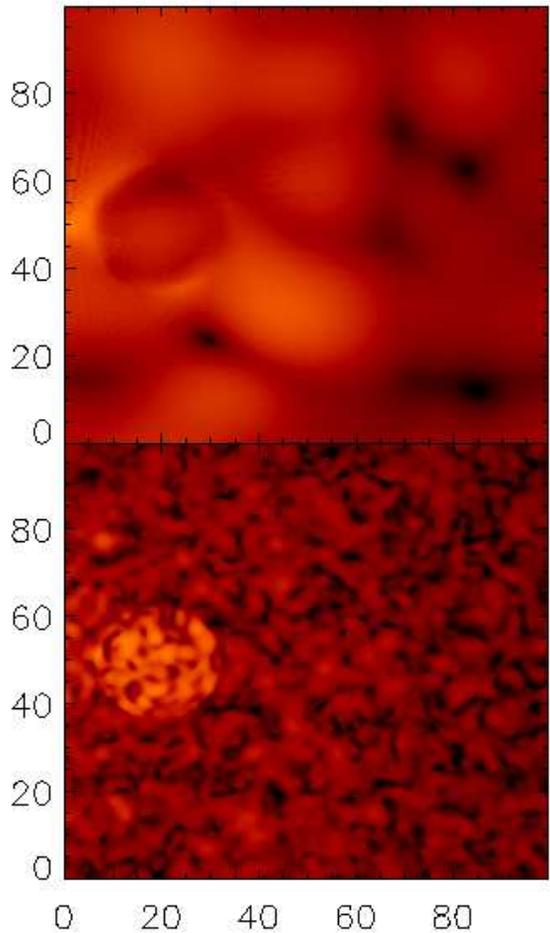}}}}
\caption{\label{}{Magnetic pressure in the plane containing the 
cluster and bubble centers. Upper and lower panels show ``draping''
and ``random'' cases, respectively. 
The coherence length in the lower panel is even smaller than that used
in the simulations to demonstrate the robustness of the method. See Appendix for more explanation.}}
\end{figure}

\begin{figure*}
\centering
\begin{minipage}[b]{\textwidth}
\centering
\includegraphics[width=1.0\textwidth, angle=0.0]{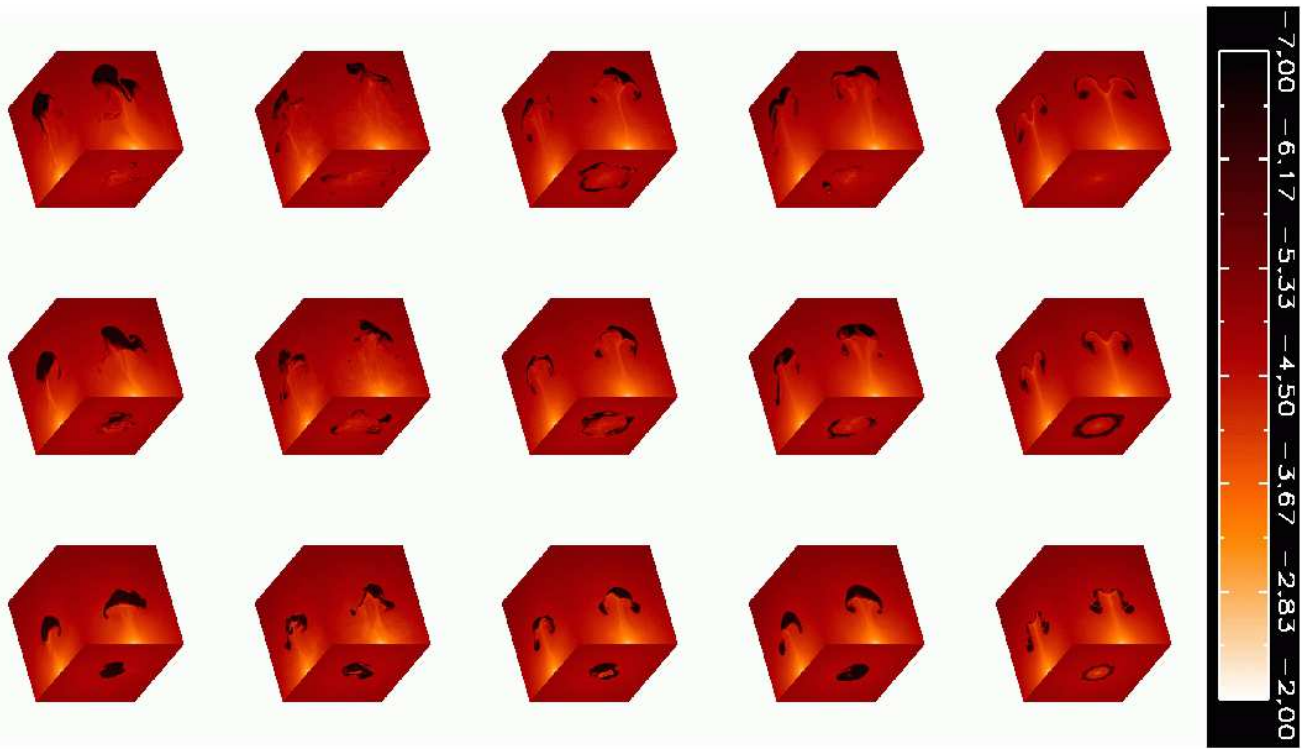}
\caption{Natural logarithm of density in the cluster. Columns show
time sequences of density for draping, random (i), random (ii), helical and
non-magnetic  
cases from left column to the right, respectively. Rows correspond to  
time of 15.0, 25.0, 35.0 code time units from bottom to top. Sides of all
boxes show densities in the planes intersecting the box and containing
its center.\label{}}
\end{minipage}
\end{figure*}

\begin{figure*}
\centering
\begin{minipage}[b]{\textwidth}
\centering
\includegraphics[width=1.0\textwidth, angle=0.0]{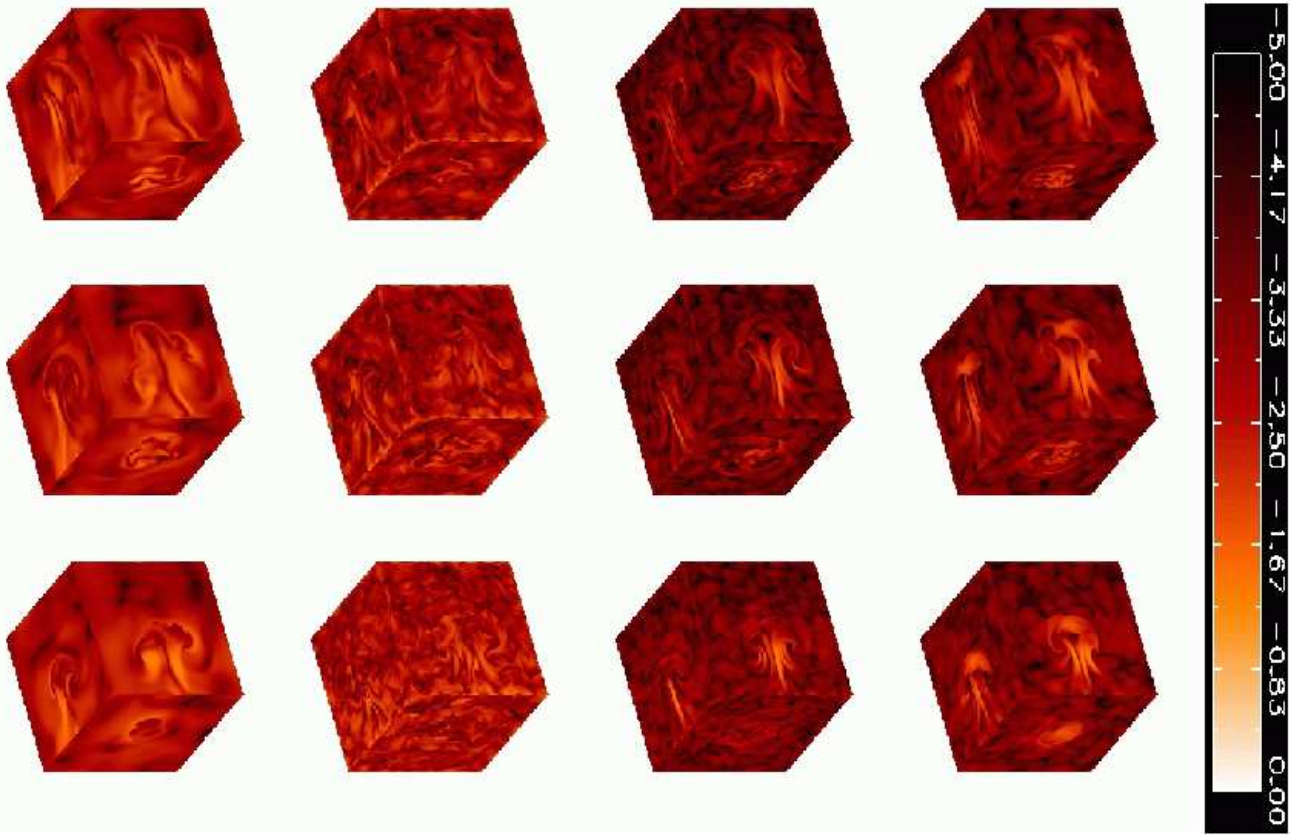}
\caption{\label{}{Magnetic pressure structure in the cluster. 
Columns show time sequences of magnetic 
pressure for draping, random (i), random (ii) and helical 
cases from left column to the right, respectively. Rows correspond to  
time of 0.0, 15.0, 25.0, 35.0 code time units from bottom to top.}}
\end{minipage}
\end{figure*}

\begin{figure*}
\centering
\begin{minipage}[b]{\textwidth}
\centering
\includegraphics[width=1.0\textwidth, angle=0.0]{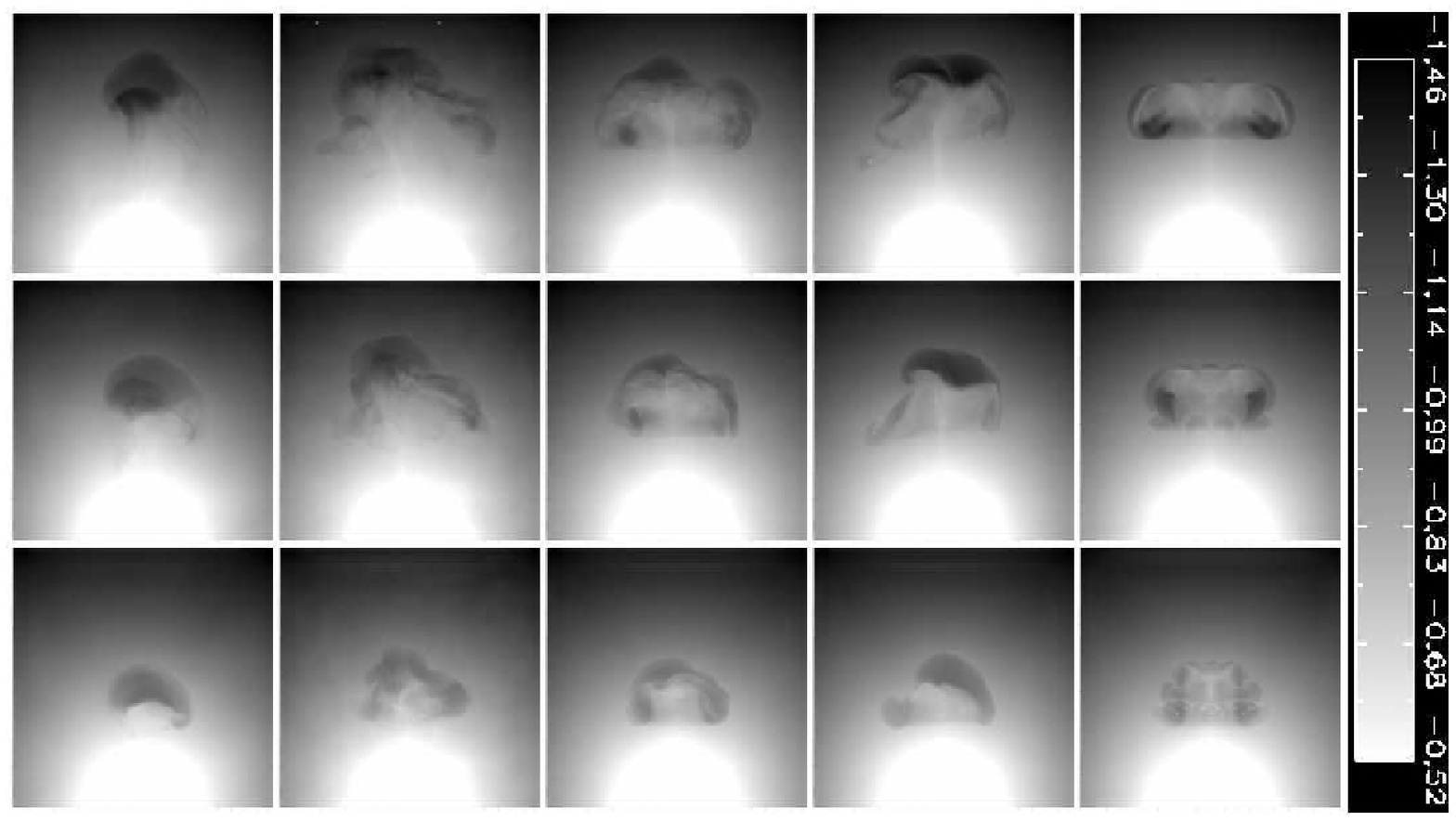}
\includegraphics[width=1.0\textwidth, angle=0.0]{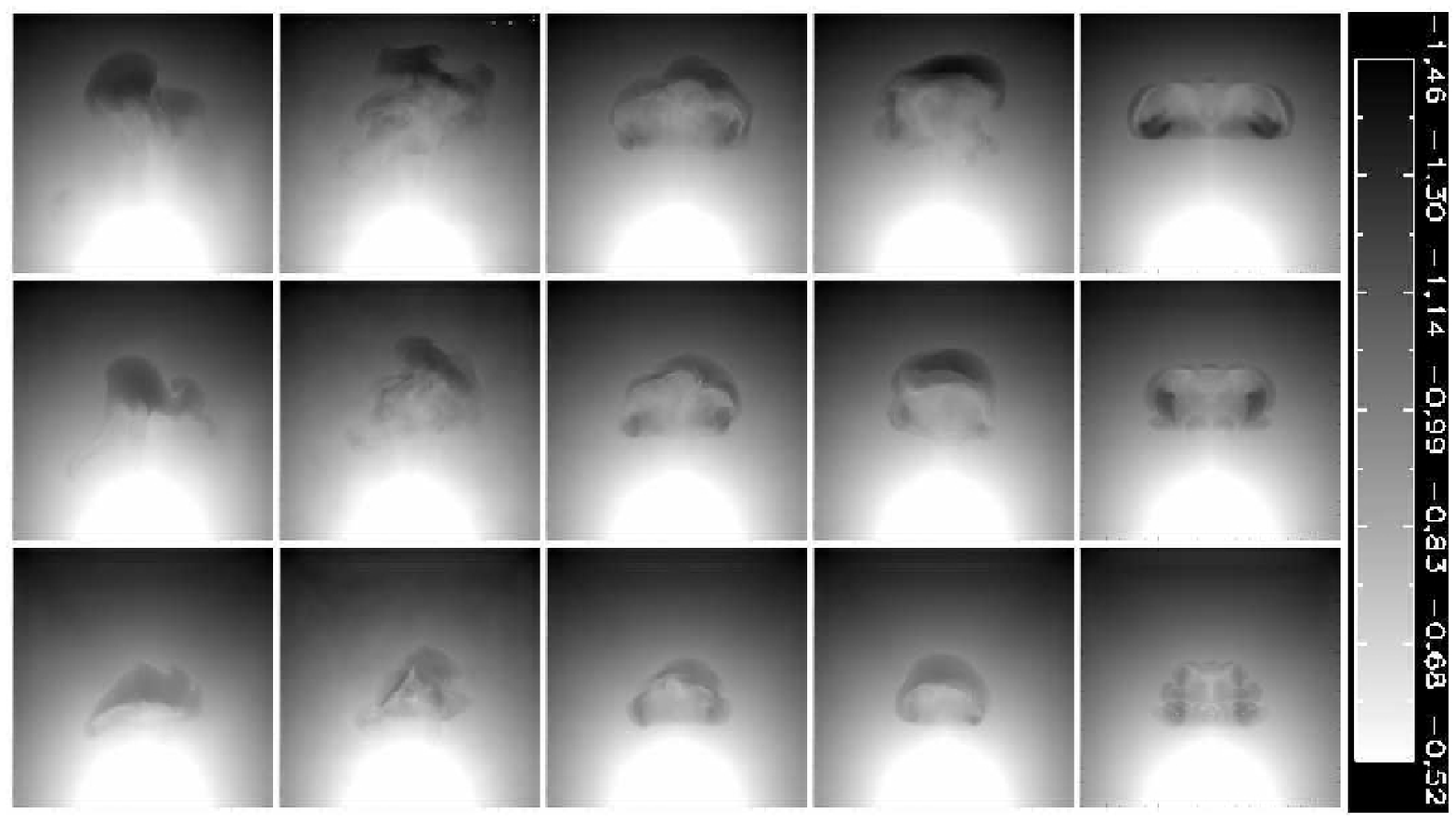}
\caption{\label{}{X-ray images of the cluster center. 
Columns show
time sequences of X-ray emission for draping, random (i), random (ii)
helical, and non-magnetic  
cases from left column to the right, respectively. Rows correspond to  
time of 15.0, 25.0, 35.0 code time units from bottom to top.
Rows 1 to 3 correspond to a different projection axis than rows 4 to 6.}}
\end{minipage}
\end{figure*}

\begin{figure*}
\centering
\begin{minipage}[b]{\textwidth}
\centering
\includegraphics[width=0.7\textwidth, angle=0.0]{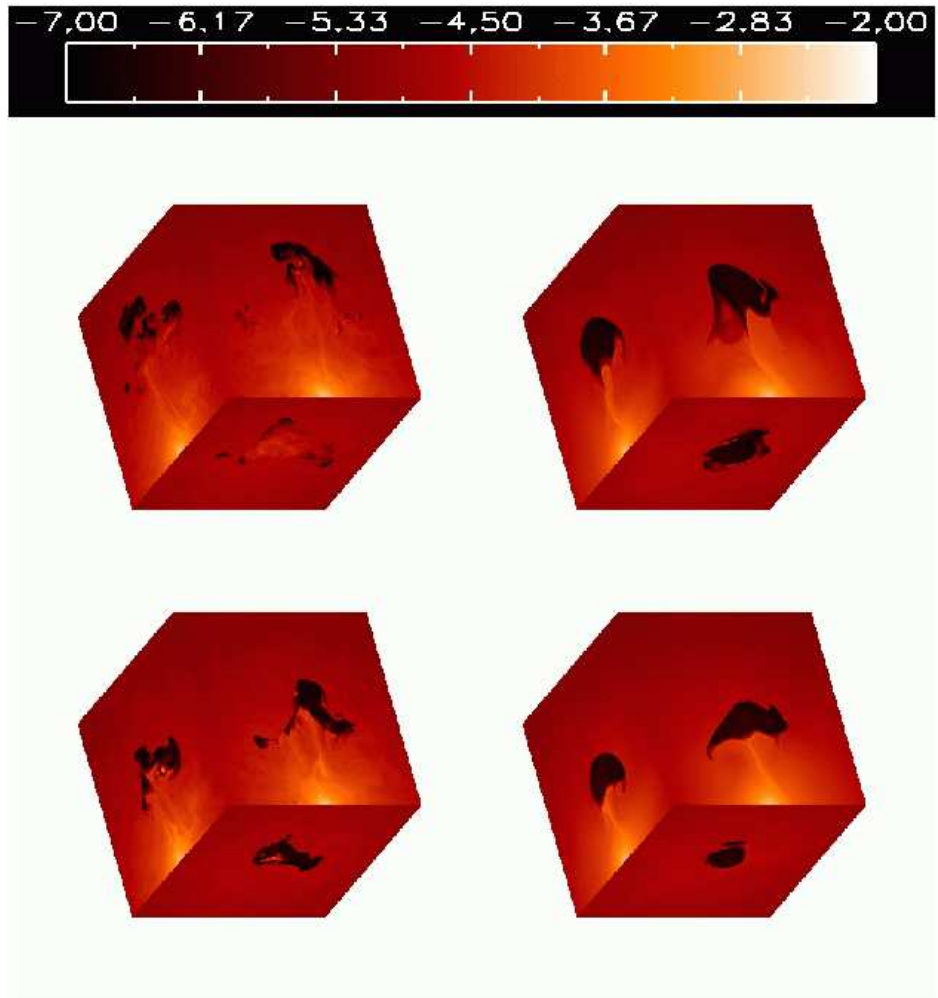}
\caption{\label{}{Natural logarithm of density distribution. 
Left column shows density for the random (i) case (lower panel t=15,
upper t = 25 code time units) but 
twice as high mean magnetic pressure.
Right column corresponds to the draping case for twice the 
mean magnetic pressure compared to the original draping case.}}
\end{minipage}
\end{figure*}

\begin{figure*}
\centering
\begin{minipage}[b]{\textwidth}
\centering
\includegraphics[width=0.7\textwidth, angle=0.0]{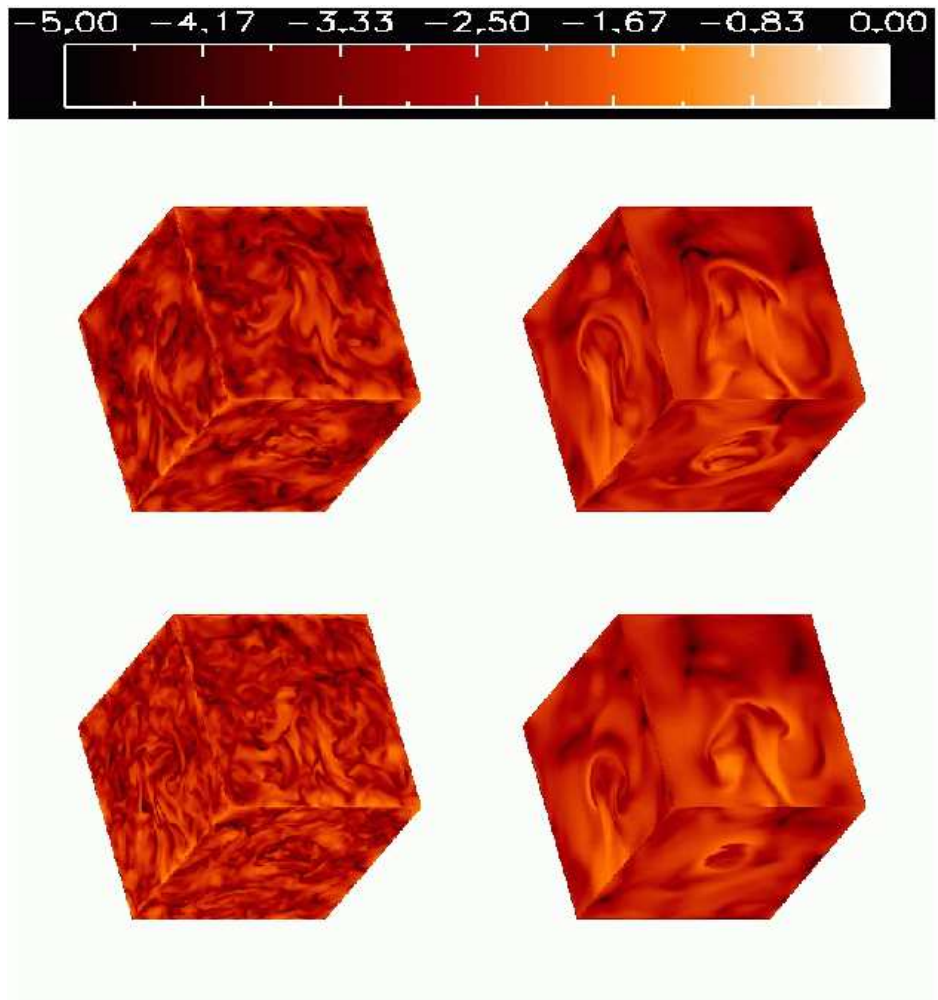}
\caption{\label{}{Same as Figure 7 but for the logarithm of magnetic pressure
distribution. Note that in the draping case (right panel), a layer of
ordered and amplified magnetic field forms on the bubble-ICM interface
and protects the bubble against disruption.}}
\end{minipage}
\end{figure*}

\begin{figure*}
\centering
\begin{minipage}[b]{\textwidth}
\centering
\includegraphics[width=0.7\textwidth, angle=0.0]{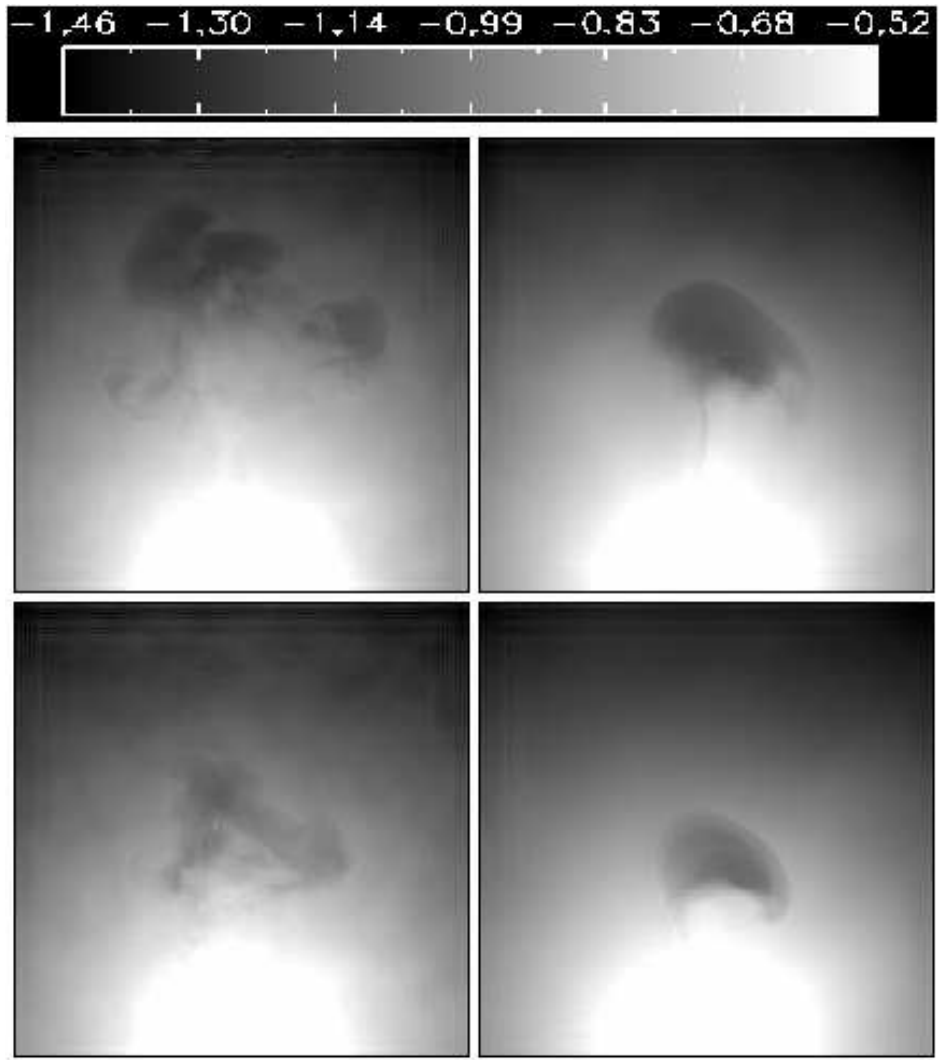}
\caption{\label{}{X-ray maps corresponding to Figures 8 and
9. Projection orientation is the same as in the upper panels in Figure 6.}}
\end{minipage}
\end{figure*}

\begin{figure}
\centering
\rotatebox{0}{\mbox{\resizebox{8.6cm}{!}{\includegraphics{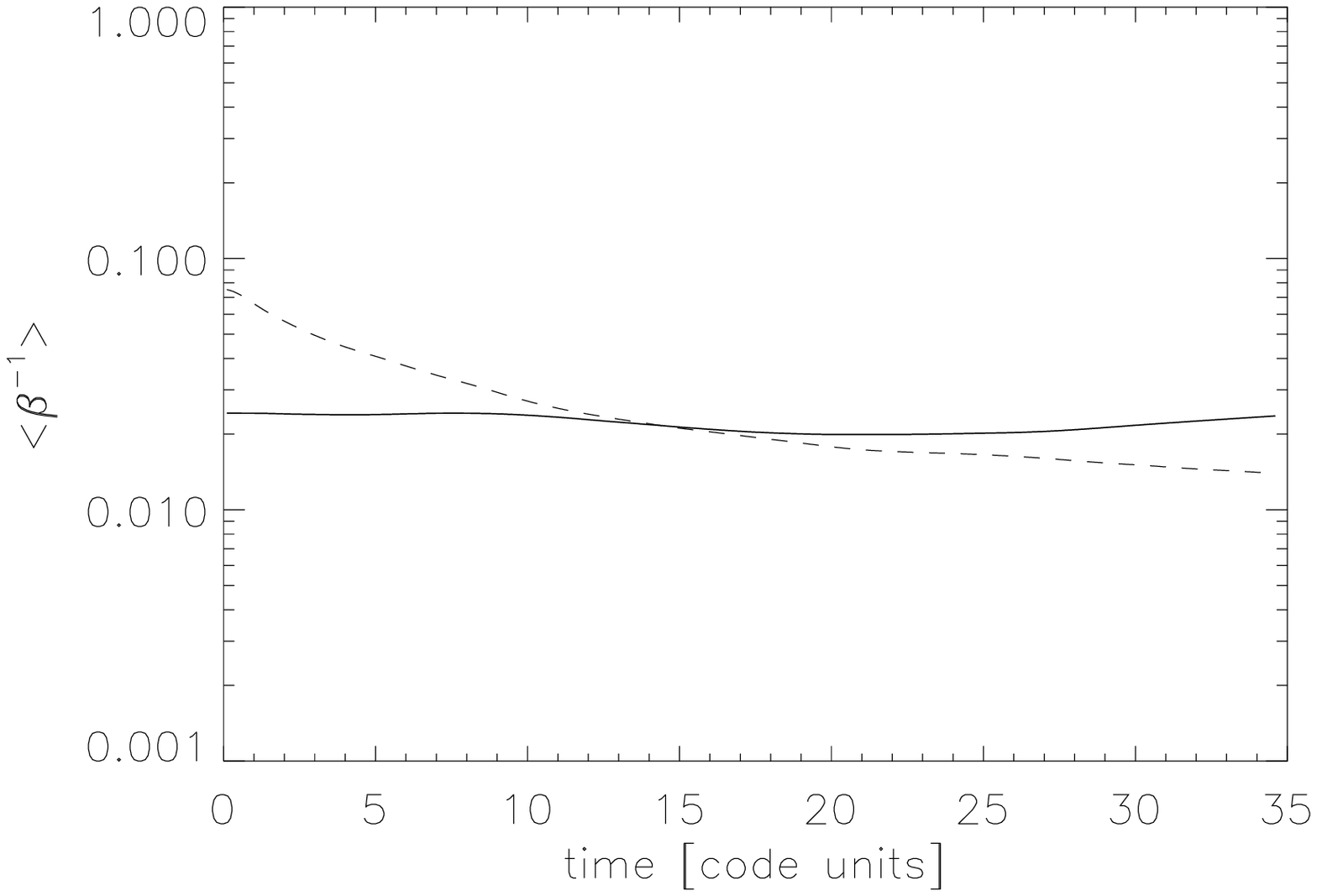}}}}
\rotatebox{0}{\mbox{\resizebox{8.6cm}{!}{\includegraphics{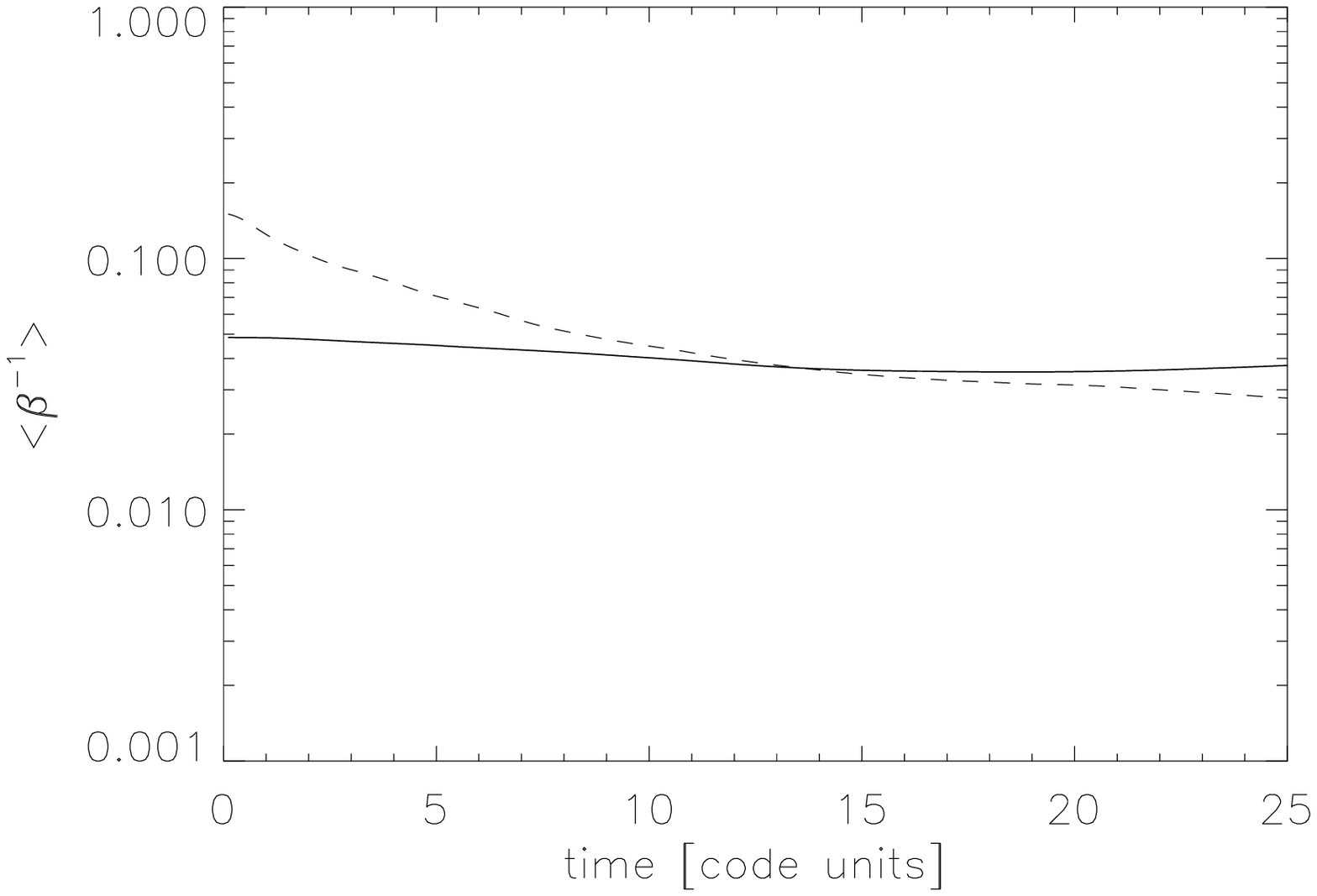}}}}
\caption{\label{}{ Evolution of the mean magnetic pressure (averaged
in the plane containing the cluster center and the initial bubble
location) compared to
gas pressure for the draping (solid lines) and random (i) cases. Lower
panel is for twice the original mean initial magnetic pressure.}}
\end{figure}

In Figure 4 we show the evolution of the gas density. Each column corresponds
to one model. Time increases from bottom to top. Snapshots were taken
at 15, 25, 35 code time units. The following five models were considered
(from left to right): 

\begin{itemize}
  \item draping model: mean $\beta\sim 40$ throughout the box
     (i.e., relative plasma $\beta$ of the bubble with respect 
     to the ICM is set to unity)
  \item random (i) model: constant mean $\beta$ throughout the box.
     The initial mean $\beta$ was chosen so as to match mean $\beta$ values 
     near $t=15$ code time units in the draping case
   \item random (ii) model : initial bubble-averaged beta inside the
     bubble the same as the mean initial $\beta$ in case random (i)
     Relative magnetic pressure inside the bubble was chosen to be 10
     times higher than the typical ambient magnetic pressure
   \item helical case: same as random (ii) but for helical fields
   \item non-magnetic case.
\end{itemize}

\indent
It is evident that the bubble morphology depends strongly on the
topology of the initial magnetic field. 
There is a striking difference between draping case (first column) 
and all remaining cases. Even though magnetic fields
are ``dynamically'' unimportant in the sense that the typical plasma
$\beta$ parameter is much greater than unity, the bubble clearly is more 
coherent in the draping case than in any other case. The cup-shaped
morphology in the draping case
in the first snapshot at 15 time units is very reminiscent of the
fossil bubble seen in the Perseus cluster. As the bubble moves up, its
shape changes and it becomes more round. The Rayleigh-Taylor
instability is prevented and so there is no evidence for strong
contamination or mixing of the bubble interior with the colder ICM material.
However, denser and colder material lifted by the rising bubble 
reverses its motion at a certain time and begins to fall back.
This leads to some stretching of the bubble in the vertical direction
in the later stages in its evolution.\\
\indent
The second column shows random (i) case. Here the gas behind the bubble
quickly penetrates its center and tends to pierce through it. 
This is evident especially when the left hand sides of the cubes 
in the first and second column are compared. The
working surface of the bubble becomes irregular and starts to
fragment. Similar behavior is observed in the random (ii) case (third
column). The two
cases differ in the strength of magnetic field with case (ii) having
stronger internal bubble fields but weaker ICM ones. Nevertheless,
there does not seem to be much qualitative difference between these 
two cases in terms of density distribution.\\
\indent
The fourth column shows the helical case. The degree of
bubble fragmentation is lower than in the random cases even though
typical magnetic field strength is similar to that of random (i)
case. Although this appears to be a weak effect, 
as explained in Section 2.2.2, helicity conservation 
should tend to stabilize the flow. This column should be compared mainly
with the third column that shows its non-helical analog.\\
\indent
The last column shows the non-magnetic case. Both Rayleigh-Taylor and
Kelvin-Helmholz instabilities develop here very quickly. The bubble is
pierced by a column of cold gas forming a ``smoke ring'' rather then 
cup-shaped or oval structure as in the draping case. The edges of
this ring are initially wrinkled by Kelvin-Helmholz instability.\\
\indent
In Figure 5 we show the distribution of magnetic pressure. As in Figure 4,
columns correspond to different models and are ordered
the same way while rows correspond to 15, 25, 35 time units from
bottom to top. This figure reveals that the reason for bubble
stability in the draping case (first column) 
is the formation of an ``umbrella'' or a
thin protective magnetic layer on the bubble working surface that
suppresses instabilities. 
Visual inspection of the figure comparing  magnetic
pressure evolution in the draping, two random, and helical cases shows
that the field geometry does not change significantly (in the average
sense) far away from the bubbles.
The upward motion of the bubble also leads to
substantial amplification and ordering 
of magnetic field in the bubble wake. This
has consequences for estimates of conduction in the ICM based on the
appearance of H$\alpha$ filaments (Fabian et al. 2003b, Hatch et al. 2006). 
It has been argued (e.g., Nipoti \& Binney 2004)
that thermal conduction has to be strongly suppressed in the ICM or
otherwise such cold filaments would be rapidly evaporated. Our result
hints at a possibility that thermal conduction may be locally 
weaker in the bubble
wake, thus preventing or slowing down filament evaporation.\\
\indent
The second column shows random (i) case. No ``umbrella'' effect seen in the
draping case is observed here. The fact that the colder gas
enters the bubble from beneath does not lead to compression of magnetic
field in the direction parallel to the bubble surface, an effect that could
prevent disruption. Even though the bubble does get disrupted,
some compression and ordering of magnetic field in the wake
occurs here just as in the draping case. In fact, this effect is
a generic feature of all the runs that we performed. The behavior of
magnetic pressure in random (ii) case, shown in the third column, is
qualitatively very similar with the difference that the magnetic wakes
are more pronounced. Even though magnetic fields get
marginally compressed near the bubble working 
surface in this case, they remain disjoint and
do not prevent the shredding of the bubble. 
\\
\indent
The last column shows magnetic pressure in the helical case.
There is a significant difference in the topology of the field inside
the bubble between this and all previous cases. Here, the upward drift of
the bubble appears to amplify magnetic field inside the bubble. Note
that while in the draping case amplification took place in a thin layer
coinciding with the working surface of the bubble, 
the amplification in the helical case is distributed 
throughout the bubble. This is
particularly obvious in the snapshot taken at 15 code time units.
We note that this means that moderate
stabilizing effect of helicity comes from fields internal to the
bubble rather than external ones as in the draping case.
No obvious amplification inside the bubble is seen in the random cases
other than the ``wake'' amplification. \\
\indent
In Figure 6 we show X-ray maps. The columns are arranged the same way as 
in Figure 4. Rows from 1 to 3 correspond to different projection
direction than rows 4 to 6 (the latter ones correspond to the viewing direction
rotated by 90 degrees around the axis intersecting the cluster center
and the original location of the bubble). 
As expected, the morphology in the draping case is very
different than in all other cases. In this case the
bubbles show up as depressions
in X-ray emissivity. The bubbles in the random case appear
irregular. They are also
brighter in their centers which is contrary to observations.
In the helical case, the bubbles seem somewhat less disturbed than in
the random case. It is possible that this case, when ``observed'' 
in synthetic {\it Chandra} data that includes instrument responses
would resemble actual bubbles more than the random case bubbles. 
The last column shows the non-magnetic case that is clearly inconsistent with
observations. Here the bubble more resembles two isolated bubbles than
a coherent feature seen in the draping case. 
We suggest that a hybrid model that combines internal helical magnetic 
fields inside the bubble with 
external draping fields may produce X-ray bubbles that closely resemble 
bubbles seen in clusters. However, different method for setting up 
initial conditions than the one considered here 
would have to be employed to model such a case while
ensuring that the initial magnetic
field configuration does not suffer from any artifacts.\\
\indent
We note that the magnetic fields in our simulations are observed to decay
with time. The decay is expected as, apart from the bubble-induced
motion, the turbulence is not continuously driven in our simulations.
As expected, the decay is faster for more tangled fields.
Even though the field decays, we note that disruption (when present) is
initiated early on in the bubble evolution. Moreover, our field
strengths are greater than those in Jones \& De Young (2005) uniform
field case and yet we do observe disruption. 
We quantified the decay in
magnetic pressure for the cases where the mean bubble and ICM $\beta$
parameters are the same (``draping'' and random (i) cases). 
In Figure 10 we show the evolution of the
mean magnetic pressure compared to gas pressure in the plane
intersecting the cluster center and the original bubble location. 
We found that, while the "random" case shows the
decay, the "draping" case is consistent with no decay.  
By construction, the mean fields in the draping case and random (i) one
approximately match at $t=15$ code time units.
The decay of the field after this time is rather slow and  magnetic field
strengths in the random and draping cases are comparable around and
after this time. Prior to $t=15$ the field in the random case exceeds
that in the draping case. It is interesting that, even though this is the
case, the random case fields lead to  bubble disruption while the
draping case shows much more coherent structures.
This demonstrates that the difference between this two cases
is primarily due to the field geometry and not its strength, at least for the
typical plasma $\beta$ considered here. Moreover, if we
would have 
considered driven turbulence then we could have afforded to start from weaker
fields in the random case and additional random motions due to
turbulence driving would be present. Both of these effects (i.e.,
weaker initial field in conjunction with additional random motions)
could only strengthen our conclusion, i.e., make the random case
bubbles fragment even more easily. We are thus conservative in
neglecting turbulence driving. We also note that our aim
was not to address the stability of the bubbles exposed to random motions.
Our objective was to discuss the effect of tangled magnetic
fields on the development of Rayleigh-Taylor and Kelvin-Helmholz
instabilities. Random
motions, be it due to turbulence driving or due to post-merger
relaxation, are inevitably going to be present in cool cluster cores
(even though cool cores tend to be more relaxed than the centers of
non-cooling flow clusters).
Their impact has to be evaluated in evaluated in sepatate studies.
Another unknown factor is how
magnetic fields are driven inside the cavities (if at all).
However, including such effects would be beyond the scope of our
investigation.

\subsection{Varying magnetic pressure}
We simulated draping and random case (i) again for exactly the same
parameters as before except for  twice as high mean magnetic pressures in both
cases. We observe the same trends with the difference that the draping
case results in slightly  more coherent bubbles while the random case
in slightly more fragmented ones. This strengthens our arguments
presented above that it is the geometry of the field rather than its
strength that is responsible for stabilizing the bubbles (at least for
the parameters considered here). In Figures 7, 8, and 9 we show
density, magnetic pressure and X-ray emissivity, respectively. 
Note that the ``umbrella'' effect mentioned above is clearly seen in the
draping case in the left column in Figure 8. This is to be contrasted
with the left hand panel corresponding to the high magnetic version of
the random (i) case where no such effect is observed. Lower
panel in Figure 10 shows the evolution of the mean magnetic pressure
compared to the gas pressure in the plane containing the center of the
cluster and the initial position of the bubble.

\section{Conclusions}

We considered three-dimensional MHD simulations of buoyant bubbles
in cluster atmospheres for varying magnetic field strengths
characterized by plasma $\beta > 1$ and for varying field
topologies. We find that field topology plays a key role in
controlling the mixing of bubbles with the surrounding ICM. 
We show that large scale 
external fields are more likely to stabilize bubbles than
internal ones but a moderate stabilizing effect due to magnetic
helicity can make internal fields play a role too.
We demonstrate that bubble morphology 
closely resembling fossil bubbles in the Perseus cluster could be realized
if the coherence of magnetic field is greater than the typical bubble size.
While it is not clear if such a ``draping'' 
case is representative of typical cluster fields, 
Vogt \& En{\ss}lin (2005) find that length scale of magnetic fields
in Hydra A is smaller than typical bubble size. 
If this also holds true in other clusters
then other mechanisms, such as viscosity, 
would be required to keep the bubbles
stable. Unfortunately, Faraday rotation method used by 
Vogt \& En{\ss}lin (2005) is not very sensitive to large scale
magnetic fields if aligned with the bubble surface.
Moreover, their maximum Likelihood method assumes a power-law relation between 
magnetic field and density, statistical isotropy for the purpose of
deprojection and a particular jet angle with respect to the line-of-sight.
Smaller angles and different magnetic field configurations might yield
a weaker decline of the power spectrum at larger scales.
Taking into account the above limitations, it is entirely possible that 
the draping case offers a viable alternative solution to 
the problem of bubble stability. 
We also suggest that a hybrid model that combines helical fields inside the 
bubble with external draping fields could be successful in explaining 
morphologies of X-ray bubbles in clusters.
Another possibility is that 
dynamically significant fields are be present inside 
the bubbles and the consequences of high-$\beta$ case 
for bubble dynamics and stability should be investigated further. 
We note that the bubbles will most
likely eventually get disrupted (partially helped by ``cosmological''
sloshing gas motions in clusters). \\
\indent
A generic feature found in our simulations is the formation of a
magnetic wake where fields are ordered and amplified. We suggest that
this effect could prevent evaporation by thermal conduction 
of cold H$\alpha$ filaments observed in the Perseus cluster.\\
\indent
The physical process of bubble mixing
in the presence of magnetic fields 
has important consequences also for modeling of mass deposition and
star formation rates in cool core clusters as well as the particle
content of bubbles and cosmic ray diffusion from them.
These issues will be further complicated by the effects of anisotropy
of transport processes due to magnetic fields. This may give rise to
the onset of magneto-thermal instability on the bubble-ICM interface
(Balbus 2004, Parrish \& Stone 2005). Studying such effects 
is beyond the scope of the
present paper but certainly deserves further investigation.

\section{Acknowledgements}
We thank the referee for a detailed and insightful report.
MR thanks Axel Brandenburg, 
Kandu Subramanian, Maxim Lyutikov, Eugene Churazov, 
Debora Sijacki, Jonathan Dursi, Maxim Markevitch, Alexei Vikhlinin
and Alexander Schekochihin for helpful discussions.
Test runs were performed on IBM p690 Regatta cluster at
Rechenzentrum Garching at the Max-Planck-Institut f{\"u}r Plasmaphysik.
Final production runs were performed on the Columbia supercomputer
at NASA NAS Ames center. It is MR's pleasure to thank 
the staff of NAS, and especially Johnny Chang and Art
Lazanoff, for their their highly professional help.
The Pencil Code community is thanked for making the code publicly
available. The main code website is located at
\texttt{http://www.nordita.dk/software/pencil-code/}. MB acknowledges
support by the Deutsche Forschungsgemeinschaft.

\bibliographystyle{mn2e}
\bibliography{mn}

\begin{thebibliography}{}



\bibitem[Balbus(2004)]{2004ApJ...616..857B} Balbus, S.~A.\ 2004, \apj, 616, 
857 

\bibitem[Blanton et al.(2003)]{2003ApJ...585..227B} Blanton, E.~L., 
Sarazin, C.~L., \& McNamara, B.~R.\ 2003, \apj, 585, 227 

\bibitem[Brandenburg et al.(2004)]{2004PhFl...16.1020B} Brandenburg, A., 
K{\"a}pyl{\"a}, P.~J., \& Mohammed, A.\ 2004, Physics of Fluids, 16,
1020 

\bibitem[Dobler et al.(2003)]{2003PhRvE..68b6304D} Dobler, W., Haugen, 
N.~E., Yousef, T.~A., \& Brandenburg, A.\ 2003, Phys. Rev. E, 68, 026304 

\bibitem[Dunn et al.(2005)]{2005MNRAS.364.1343D} Dunn, R.~J.~H., Fabian, 
A.~C., \& Taylor, G.~B.\ 2005, \mnras, 364, 1343 

\bibitem[Dunn \& Fabian(2004)]{2004MNRAS.355..862D} Dunn, R.~J.~H., \& 
Fabian, A.~C.\ 2004, \mnras, 355, 862 

\bibitem[Dunn \& Fabian(2004)]{2004MNRAS.355..862D} Dunn, R.~J.~H. 2006,
PhD Thesis, Univ. of Cambridge

\bibitem[En{\ss}lin(2003)]{2003A&A...401..499E} En{\ss}lin, T.~A.\ 2003, 
\aap, 401, 499 

\bibitem[Vogt \& En{\ss}lin(2005)]{2005A&A...434...67V} Vogt, C., \& 
En{\ss}lin, T.~A.\ 2005, \aap, 434, 67 

\bibitem[Fabian et al.(2003a)]{2003MNRAS.344L..43F} Fabian, A.~C., Sanders, 
J.~S., Allen, S.~W., Crawford, C.~S., Iwasawa, K., Johnstone, R.~M., 
Schmidt, R.~W., \& Taylor, G.~B.\ 2003, \mnras, 344, L43 

\bibitem[Fabian et al.(2003b)]{2003MNRAS.344L..48F} Fabian, A.~C., Sanders, 
J.~S., Crawford, C.~S., Conselice, C.~J., Gallagher, J.~S., \& Wyse, 
R.~F.~G.\ 2003, \mnras, 344, L48 

\bibitem[Fabian et al.(2006)]{2006MNRAS.366..417F} Fabian, A.~C., Sanders, 
J.~S., Taylor, G.~B., Allen, S.~W., Crawford, C.~S., Johnstone, R.~M., \& 
Iwasawa, K.\ 2006, \mnras, 366, 417 

\bibitem[Fryxell et al.(2000)]{Fryxell et al.(2000)}Fryxell et
al. 2000, ApJS, 131, 273

\bibitem[Hatch et al.(2006)]{2006MNRAS.367..433H} Hatch, N.~A., Crawford, 
C.~S., Johnstone, R.~M., \& Fabian, A.~C.\ 2006, \mnras, 367, 433 

\bibitem[Haugen et al.(2003)]{2003ApJ...597L.141H} Haugen, N.~E.~L., 
Brandenburg, A., \& Dobler, W.\ 2003, \apjl, 597, L141 

\bibitem[Haugen et al.(2004)]{2004MNRAS.353..947H} Haugen, N.~E.~L., 
Brandenburg, A., \& Mee, A.~J.\ 2004, \mnras, 353, 947 

\bibitem[Heinz \& Churazov(2005)]{2005ApJ...634L.141H} Heinz, S., \& 
Churazov, E.\ 2005, \apjl, 634, L141 

\bibitem[Jones \& De Young(2005)]{2005ApJ...624..586J} Jones, T.~W., \& De 
Young, D.~S.\ 2005, \apj, 624, 586 

\bibitem[Kaiser et al.(2005)]{2005MNRAS.359..493K} Kaiser, C.~R., 
Pavlovski, G., Pope, E.~C.~D., \& Fangohr, H.\ 2005, \mnras, 359, 493

\bibitem[Lazarian(2006)]{2006ApJ...645L..25L} Lazarian, A.\ 2006, \apjl, 
645, L25 

\bibitem[Lyutikov(2006)]{2006MNRAS...373..73} Lyutikov, M. 2006, MNRAS, 373, 73

\bibitem[Nakamura et al.(2006)]{2006astro.ph..9007N} Nakamura, M., Li, H., 
\& Li, S.\ 2006, ArXiv Astrophysics e-prints, arXiv:astro-ph/0609007 

\bibitem[Nipoti \& Binney(2004)]{2004MNRAS.349.1509N} Nipoti, C., \& 
Binney, J.\ 2004, \mnras, 349, 1509 

\bibitem[Parrish \& Stone(2005)]{2005ApJ...633..334P} Parrish, I.~J., \& 
Stone, J.~M.\ 2005, \apj, 633, 334 

\bibitem[Pizzolato \& Soker(2006)]{2006MNRAS.371.1835P} Pizzolato, F., \& 
Soker, N.\ 2006, \mnras, 371, 1835 

\bibitem[Reynolds et al.(2005)]{2005MNRAS.357..242R} Reynolds, C.~S., 
McKernan, B., Fabian, A.~C., Stone, J.~M., \& Vernaleo, J.~C.\ 2005, 
\mnras, 357, 242 

\bibitem[Robinson et al.(2004)]{2004ApJ...601..621R} Robinson, K., et al.\ 
2004, \apj, 601, 621 

\bibitem[Ruszkowski \& Begelman(2002)]{2002ApJ...573..485R} Ruszkowski, M., 
\& Begelman, M.~C.\ 2002, \apj, 573, 485 

\bibitem[Ruszkowski et al.(2004)]{2004ApJ...615..675R} Ruszkowski, M., 
Br{\"u}ggen, M., \& Begelman, M.~C.\ 2004, \apj, 615, 675 

\bibitem[Ruszkowski et al.(2004)]{2004ApJ...611..158R} Ruszkowski, M., 
Br{\"u}ggen, M., \& Begelman, M.~C.\ 2004, \apj, 611, 158 

\bibitem[Schekochihin \& Cowley(2006)]{SC(2006)}
Schekochihin, A., \& Cowley, S.~C.\ 2006, Physics of Plasmas, 13, 056501

\end{thebibliography}

\section{Appendix}

Initial magnetic fields were computed outside the main code
and the setup was performed in two stages. In the first phase,
we generated stochastic fields by three-dimensional 
inverse Fourier transform (FFT) of magnetic
field that in ${\bf k}$-space had the amplitude given by

\begin{equation}
B\propto k^{-11/6}\exp(-(k/k_{1})^4)\exp(-k_{2}/k),
\end{equation}

\noindent
where $k=(k_{x}^2+k_{y}^{2}+k_{z}^2)^{1/2}$, and
$k_{1} = 2\pi/dx$, $k_{2} = 0.3$, $dx = x_{\rm box}/32$, 
$x_{\rm box} = 10^{2}$ for helical or random cases 
and $k_{1} = 2\pi/(2r_{\rm bub})$ and 
$k_{2} = 0.05$ in the draping case, where $x_{\rm box}$ is the size of
the computational box in code units. (see below)\\
\indent
All three components of magnetic field were treated
independently which ensured that the final distribution of ${\bf
B}({\bf r})$ had random phase. That is, for example for the x
component of the magnetic field, we set up a complex field
such that 

\begin{equation}
({\rm Re}({B_{x}}({\bf k})),{\rm Im}({B_{x}}({\bf k})))=(G(u_{1})B,G(u_{2})B), 
\end{equation}

\noindent
where $G$ is a function of a uniform random 
deviate $u_{1}$ or $u_{2}$ that returns Gaussian-distributed values.
For vanishing exponential cutoff terms, the above prescription would 
give classical Kolmogorov turbulence spectrum. 
Whereas there is no generally accepted 
justification for magnetic spectrum to have
a Kolmogorov distribution, our parameter choice for the ``random'' case
resembles that seen in the Hydra cluster (Vogt \& En{\ss}lin 2005).
One-dimensional energy power spectra $kE(k)$ [erg s$^{-1}$ cm$^{-3}]$
of magnetic field fluctuations are shown in Figure 2. \\
\indent
After the initial field has been set up in k-space, 
we implement a magnetically
isolated bubble. This phase is performed according to the following
iterative scheme:\\

\indent
(1) divergence cleaning in ${\bf k}$-space:

\begin{equation}
{\bf B}({\bf k})\longmapsto ({\bf 1}-\hat{\bf k}\hat{\bf k}){\bf B}({\bf k}), 
\end{equation}

\noindent
where $\hat{\bf k}$ is a unit vector in ${\bf k}$-space. In the case
of helical fields we also act on the field with helicity operator
defined as:

\begin{equation}
{\bf B}({\bf k})\longmapsto 
\frac{{\bf 1}+\alpha i\hat{\bf k}\times}{(1+\alpha^2)^{1/2}}{\bf B}({\bf k}),
\end{equation}

\noindent
where $\alpha = 1$ (maximum helicity case).
The helicity step is only performed in the first loop of the iteration.
Note that imposing helicity on the field does not change the power
spectrum of magnetic energy fluctuations.

\indent
(2) inverse FFT of the new ${\bf B(k)}$ to real space. Each component
of ${\bf B(k)}$ is independently acted upon with a three-dimensional
inverse FFT.

\indent
(3) applying a projection operator to isolate the bubble magnetically. 
The projection operation modifies the field in the following way

\begin{equation}
{\bf B}({\bf r})\longmapsto[{\bf 1}-g(r)\hat{\bf r}
\hat{\bf r}]{\bf B}({\bf r}), 
\end{equation}

\noindent
where $\hat{\bf r}$ is a unit
vector in real space, $r$ is the distance from the bubble center, 
and $g(r)=1-|\cos[0.5\pi(x+\Delta x-1)/\Delta x]|$
for $1-\Delta x < x < 1+\Delta x$ and 
$g(r)=0$ otherwise and $\Delta x=0.25$, where $x=r/r_{\rm bub}$. 
Note that both the field just inside and outside the bubble
are acted upon by this operator. Applying this operator
results in a field that possesses some divergence.

(4) changing relative magnetic
pressure inside the bubble (done only during the
first loop of the iteration process) 
Plasma $\beta$ parameter is
given by $\beta(r) = \beta_{\rm rel}g_{b}(r)+1-g_{r}(r)$, 
where $\beta_{\rm rel}$
is the relative plasma $\beta$ parameter between the ICM and the
bubble. The $g_{b}(r)$ term is given by 
$g_{b}(r)=1+\cos[0.5\pi (x+\Delta_{b} x-1)/\Delta_{b} x]$,
where $x=r/r_{\rm bub}$ for $1 < x < 1+\Delta_{b} x$ 
and $g(r)=0$ otherwise. 
In the above expression, $r_{\rm bub}$ is the bubble
radius and $\Delta_{b} x=0.15$.

(5) computing the FFT of ${\bf B}({\bf r})$ 
and going back to (1). Iterations are
    performed until the following convergence criterion is met:
$\int (\nabla\cdot {\bf B}({\bf r}))^2 dV dx^2/\int B^{2}dV<10^{-4}$, where
    integrations are performed for $|x-1|<\Delta x$.\\

\indent
At this point a convergent and divergence-free ${\bf B}({\bf r})$ 
field has been set up.
Inverting such field to obtain the vector potential ${\bf A}({\bf r})$ 
does not result in any loss of information.
Note that inverting {\bf B} to get {\bf A} before completing the 
iteration process would result in a vector potential that
(by definition) would give divergence-free magnetic field but the
bubble would not be magnetically isolated. \\
\indent
The equation
${\bf B}=\nabla\times{\bf A}$ is then solved for ${\bf A}({\bf r})$
using a variant of a spectral method as follows.\\

\indent
(1) we Fourier transform ${\bf B}=\nabla\times{\bf A}$\\
\indent
(2) keeping ${\bf B(k)}$ and ${\bf k}$ constant we rotate 
the k-space coordinate
    system first around the $k_{z}$-axis until the projection of
the ${\bf k}$ vector on the $(k_{x},k_{y})$ plane coincides with 
$k_{x}$ axis and then around $k_{y'}$-axis until $k_{z'}$ coincides
with the ${\bf k}$ vector. This way the
$k_{z''}$-axis is aligned with the ${\bf k}$ vector. That is,

\begin{equation}
{\bf B^{''}}({\bf k^{''}})\longmapsto R(y',\alpha_{y'})R(z,\alpha_{z}){\bf B}({\bf k}),
\end{equation}

\noindent
where $R(i,\alpha_{i})$ are rotation matrixes around axis $i$ by angle
$\alpha_{i}$.

\indent
(3) we invert ${\bf B}''=-i{\bf k}''\times{\bf A}''$ 
(which, in this frame of reference, is trivial as 
${\bf k^{''}}=(0,0,1)$).

(4) we rotate ${\bf A}''$ back and in reverse order, i.e., 

\begin{equation}
{\bf A^{''}}({\bf k^{''}})\longmapsto R(z,-\alpha_{z})R(y',-\alpha_{y'}){\bf A^{''}}({\bf k^{''}}),
\end{equation}

\noindent
where ${\bf A}({\bf k})$ is the Fourier transform of the required 
vector potential. Finally, 
we inverse FFT ${\bf A}({\bf k})$ to obtain the vector potential 
in real space. \\

\indent
The code uses the vector potential as its ``magnetic'' variable which
ensures that divergence of magnetic field is strictly zero throughout
the simulation. The reason our initial conditions were not set up
    directly in terms of the vector potential is that adding bubble as a
    distortion in the fluctuating background vector potential leads to
    significant gradients at the boundary between the bubble and the
    ICM. These gradients translate into very strong artificial enhancements
    in the magnetic field surrounding the bubble as we have seen in our
    experiments with that kind of setup.\\
\indent
We also note that the field
set up in this way is not force-free. It is not possible to set up
    force-free field for isotropic turbulence case due to mode
    coupling. However, initial imbalance in magnetic forces are small
    compared to the buoyancy force acting on bubbles. Moreover,
    realistic turbulence is not expected to be force-free in any case 
    as it has to be continuously driven to prevent its decay.\\
\indent
Note that the $\beta$ parameter and magnetic field strength 
obtained using the above method would 
have arbitrary overall normalization.
The actual normalization of the magnetic flux is obtained by demanding
that the $\beta$ parameter has a certain value inside the bubble, i.e.,
${\bf B}({\bf r})\longmapsto {\bf B}({\bf r})(P_{\rm gas}\beta_{\rm bub}^{-1}/
\langle P_{B}\rangle_{\rm bub})^{1/2}$, 
where $P_{B}=B^{2}/2\mu_{o}$ is magnetic
pressure, $P_{\rm gas}$ is the gas pressure, $\beta_{\rm bub}$ is 
the required plasma $\beta$ inside the bubble, and averaging is done
over the bubble volume.
We note that this method works best when applied to high-$\beta$ cases
as then the imbalance between the total pressure in the bubble and the ICM 
is smallest.\\
\indent
The final form of our initial conditions for the distribution of
magnetic pressure for the draping and random cases is shown in Figure
3. This figure shows that our method for generating initial conditions
does not produce any spurious features on the bubble/ICM interface.
The coherence length in the lower panel is even smaller than that used
in the simulations. This has been done to demonstrate robustness of
the method. The map shows natural logarithm of magnetic pressure 
in arbitrary units.\\
\indent
Note that magnetic field configurations were generated from the same
random seed, which means that, despite the differences due to
different power spectra, $\beta$ values, helicities, etc., the fields
were as similar as possible. This permits a better comparison of the
consequences of the mentioned differences.

\label{lastpage}
\end{document}